\documentclass[a4paper]{jpconf}
\usepackage{graphicx}

\renewcommand{\pt}{$p_\mathrm{T}$}
\newcommand{\meanpt}{$\langle p_\mathrm{T}\rangle$}
\newcommand{\Ncoll}{N$_{\mathrm{coll}}$}
\newcommand{\Npart}{N$_{\mathrm{part}}$}
\newcommand{\Tcent}{T$_{\mathrm{central}}$}

\begin{document}
\title{Modeling Quark Gluon Plasma  Using CHIMERA}
\author{Betty Abelev}
\address{ Physics Division, Lawrence Livermore National Laboratory, 7000 East Avenue, L-211, Livermore, CA 94550-9698}
\ead{abelev@llnl.gov}

\begin{abstract} We attempt to model Quark Gluon Plasma (QGP) evolution from the initial Heavy Ion collision to the final hadronic gas state by combining the Glauber model initial state conditions with eccentricity fluctuations, pre-equilibrium flow, UVH2+1 viscous hydrodynamics with lattice QCD Equation of State (EoS), a modified Cooper-Frye freeze-out and the UrQMD hadronic cascade.  We then evaluate the model parameters using a comprehensive analytical framework which together with the described model we call CHIMERA.    Within our framework, the initial state parameters, such as the initial temperature (T$_{\mathrm{init}}$), presence or absence of initial flow, viscosity over entropy density ($\eta$/s) and different Equations of State (EoS), are varied and then compared simultaneously to several experimental data observables: HBT radii, particle spectra and particle flow. $\chi^2$/nds values from comparison to the experimental data for each set of initial parameters will then used to find the optimal description of the QGP with parameters that are difficult to obtain experimentally, but are crucial to understanding of the matter produced. 
\end{abstract}

\section{Introduction}

Recreating Quark Gluon Plasma (QGP), matter that existed in the first millisecond after the Big Bang, has been the goal of experiments at RHIC and now at the LHC.  QGP is created during the course of a collision of massive relativistic ions, such as gold ions in the Brookhaven Laboratory or lead ions at CERN, and has been shown to exhibit the properties of a viscous liquid \cite{Perfect LiquidSTAR}\cite{Perfect_LiquidPHENIX}.  To understand the properties of the matter created in these collisions, the evolution of matter during the collision itself must be understood. Currently, we understand the basic stages of matter evolution, but lack the nuanced understanding of interdependencies and variation in parameters required for precise description of the collision.   

Our present understanding of a heavy ion collision can be summarized as follows.  After collision time $\tau\approx1$ fm/c, a strongly interacting liquid of quarks and gluons is formed, which is described with various accuracy by hydrodynamical models.  As the gas cools, mesons and hadrons ``freeze out" and start interacting as colorless objects, until a phase of a relatively stable gas of hadrons is reached.  

In this work we attempt to evaluate numerically the validity of models describing this picture by systematically varying model parameters and calculating $\chi^2$ distributions of proximity of fit to the experimental data.  The evaluating algorithm used for these purposes is called CHIMERA, which stands for Comprehensive Hydrodynamical Integrated Modelling Evaluation and Reporting Algorithm, and will be described in detail in the next section.

\section{Model description}

\begin{figure}
\begin{center}
\includegraphics[width=5.50 in]{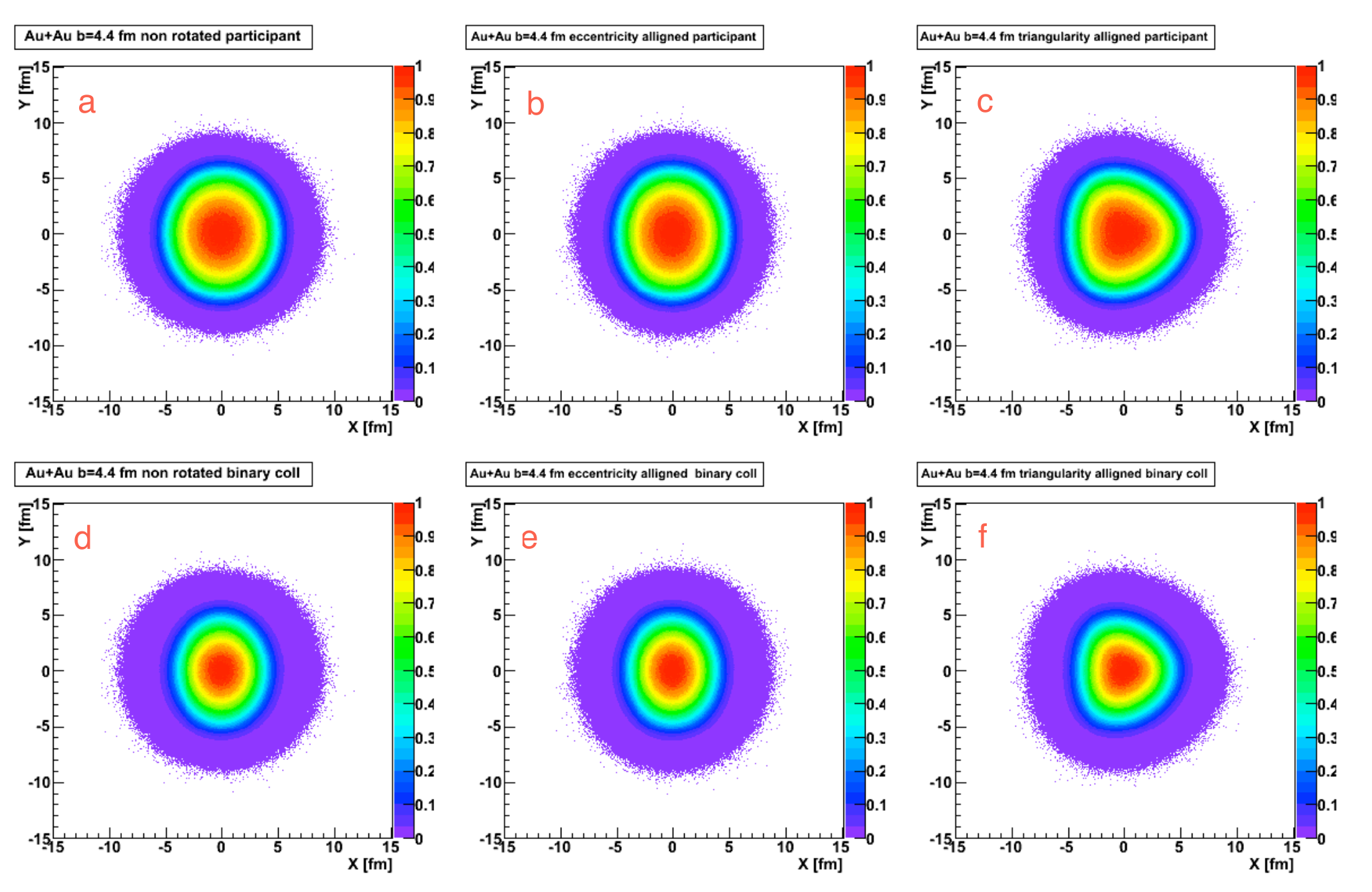}
\caption{\label{figure1} Hydrodynamical initial conditions energy density distributions for Au+Au collisions at $\sqrt{s_{NN}}=200$ GeV with $b=$4.4 fm, scaled as \Npart\ ($a,b,c$), or \Ncoll\ ($d,e,f$).  Panels $a$ and $d$ were produced using optical Glauber geometry, panels $b$ and $c$ using Monte Carlo Glauber, and panels $c$ and $f$ using Glauber geometry with event-by-event fluctuations.}
\end{center}	
\end{figure}

\begin{figure}
\begin{center}
\includegraphics[width=4.50 in]{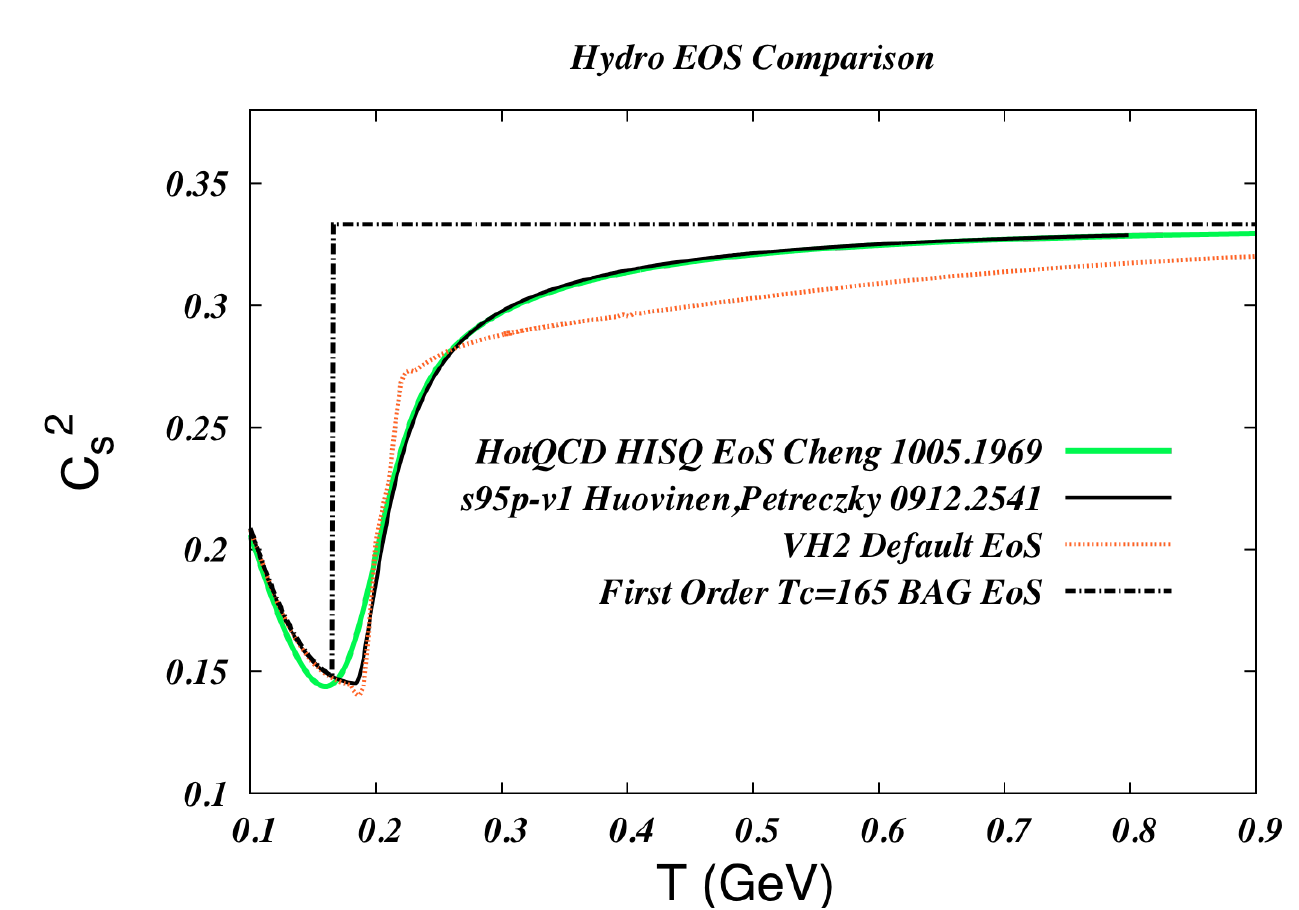}
\caption{\label{figure2} Equation of state functions from HotQCD  (solid green curve), s95p-v1 (solid black curve), VH2 lattice-inspired EoS (pink curve), and the MIT Bag model EoS with a first order transition.}
\end{center}	
\end{figure}

CHIMERA is hydrodynamics-based framework, thus we first start by determining initial inputs and an initial energy distribution.  In the analysis presented here, the initial conditions are computed using Monte Carlo Glauber model of the colliding nuclei \cite{Glauber}   (e.g., Figure \ref{figure1}, panels $b$ and $e$).    However, CHIMERA is capable of handling several initial condition implementations, including optical Glauber initial conditions (Figure \ref{figure1}, panels $a$ and $d$), Glauber conditions with fluctuations (Figure \ref{figure1}, panels $c$ and $f$), and Colour Glass Condensate-inspired \cite{CGC} initial conditions (not shown).  We implement initial conditions with either the number of collisions (\Ncoll) or number of participants (\Npart) scaling. In the future, we plan to introduce a dial-up mechanism for this scaling, so that the initial conditions would scale with a predetermined mix of  \Ncoll\ and \Npart, as is closer to experimental observations. The Equation of State (EoS) used as a CHIMERA input is obtained from Lattice QCD calculations by the HotQCD collaboration \cite{HotQCD}, and agrees well with s95p-vi EoS by Huovinen and Petreczky \cite{Petrecky}.  This EoS is much softer than the resonance gas EoS  predicted by the MIT Bag Model \cite{MITBag}, but is stiffer than the lattice-inspired EoS used in VH2 Hydro \cite{Hydro}. All four equations of state are shown in Figure \ref{figure2}.  

As the last step before starting viscous hydrodynamics, we employ pre-equilibrium flow, i.e., flow that originates from the non-equilibrated matter at early times (approximately before $\tau=1$ fm/c), to start  the evolution of the heavy ion collision.  This is motivated by fits to variables describing the collective behaviour of the fireball, such as Hanbury Brown-Twiss (HBT) radii and collective flow parameters. Pre-equilibrium flow is universal when describing a boost-invariant system with a traceless stress-energy tensor \cite{Pratt}.  Since our model utilizes a 2+1 dimensional hydrodynamic evolution (see below), we are well-justified in employing pre-equilibrium flow in our description of heavy ion collisions.

Using the set up described above, we run VH2 2D+1 viscous hydrodynamics developed by Luzum and Romatschke \cite{Hydro} on the energy-density distribution that resulted from the above assumptions.  At the end of the hydrodynamic evolution, we employ Cooper-Frye freezeout \cite{CooperFrye} with viscous corrections, and then start the UrQMD hadronic cascade \cite{Bass} to obtain final particle distributions.  

After ensuring that the resultant yields and mean \pt\ (\meanpt) of the produced particles match those of experimental data, we employ our analytical framework to draw numerical comparisons between our model results and those from experiment.  This is done by describing our modelling data using Chebyshev polynomials, and then calculating the $\chi^2$/ndf between these polynomials and experimental data.  The $\chi^2$ is calculated taking into account point-to-point correlated, uncorrelated, and scaling uncertainties (A, B, and C-type errors, as described in \cite{PhenixPi}).  Performing this fit and calculation for a grid of initial state parameters, such as viscosity over entropy-density ($\eta/s$) and the maximum temperature of the evolving medium (T$_{\mathrm{central}}$), we are then able to do a global comparison of $\chi^2$ minima for a variety of soft observables: particle spectra, elliptic flow coefficients, and femtoscopic measurements.  

\section{Data Analysis}

As mentioned in the previous section, randomized Glauber initial conditions for Au+Au collisions at $\sqrt{s_{NN}}=200$ GeV with an impact parameter ($b$) of 4.4 fm, shown in panels $b$ and $e$ of Figure \ref{figure2}, are used as an input for this study.  The main goal of the analysis is two-fold: first, to study the differences due to inclusion or omission of pre-equilibrium flow.  Second, to study the impact of using \Ncoll\ vs. \Npart\ scaling on the accuracy of describing experimental data. Four sets of modelling data were generated: two sets with pre-equilibrium flow turned on (one scaled with \Ncoll, the other with \Npart), and two sets with pre-equilibrium flow turned off (also one with \Ncoll\ scaling and the other with \Npart\ scaling).  We varied T$_{\mathrm{central}}$ (highest temperature of the medium) from 320 to 360 MeV, and $\eta/$s from 0.001 to 0.36.  The VH2 EoS was replaced by HotQCD HISQFIX EoS \cite{HotQCD}.   Cooper-Fry freeze-out was started when the temperature of the medium reached 165 MeV (T$_{\mathrm{freeze-out}}$), roughly corresponding to the temperature of chemical freeze-out derived from the thermal model \cite{Becattini}.  UrQMD was started at T$_{\mathrm{switch}}=140$ MeV.

\begin{figure}
\begin{center}
\includegraphics[width=5.50 in]{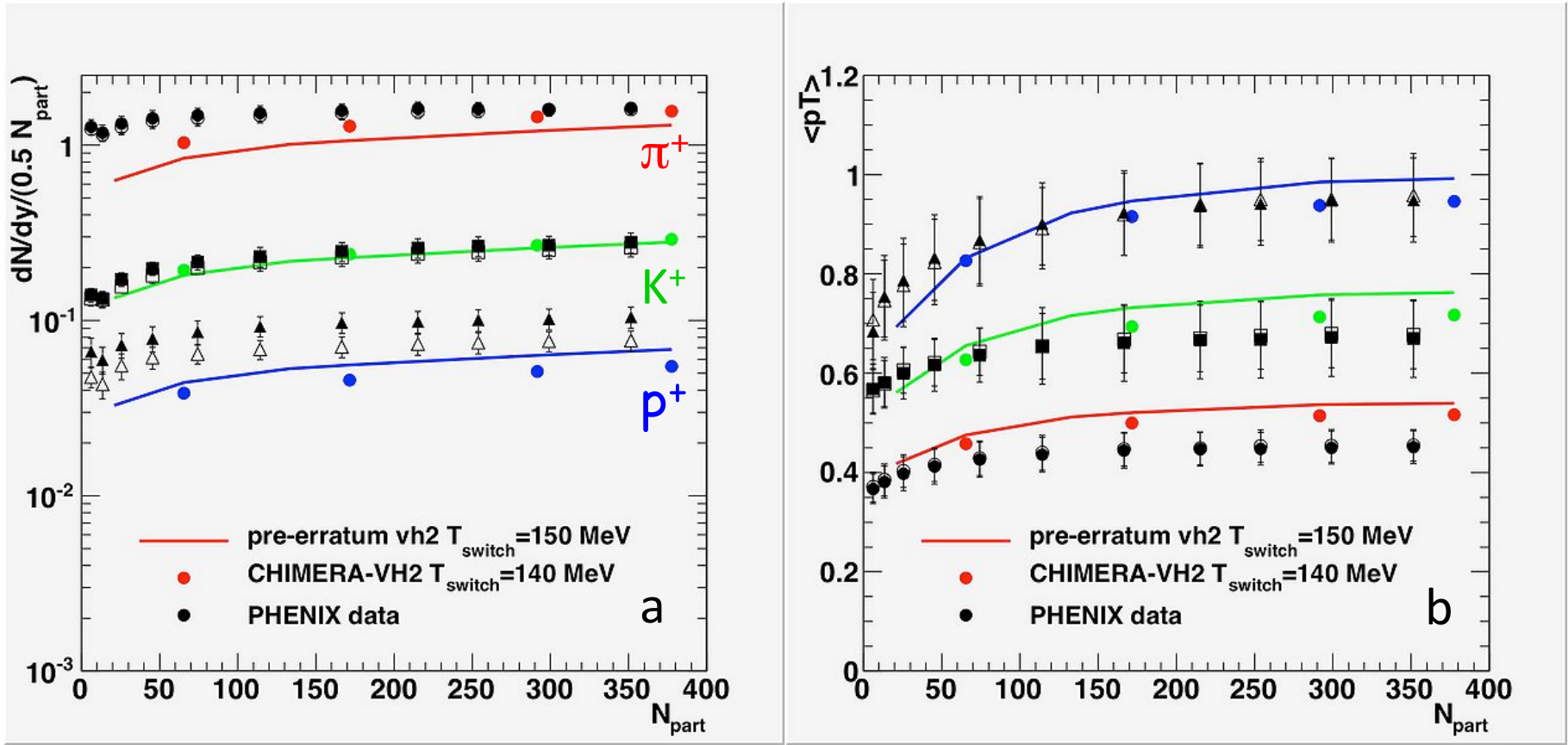}
\caption{\label{figure3} $\pi^+$, K$^+$, and p$^+$ dN/dy (panel $a$) and \meanpt\ (panel $b$) distributions from $\sqrt{s_{NN}}=200$ GeV 10-20\% central Au+Au collisions, modelled using CHIMERA (collored filled symbols), VH2 pre-errata data with T$_{\mathrm{switch}}$=150 (color curves), and experimental data from PHENIX (black and open symbols).}
\end{center}
\end{figure}

To check for consistency, we ran a sample of Au+Au 200 GeV data without any centrality selections.  We calculate yields per unit of rapidity (dN/dy) and  \meanpt\ as a function of \Npart\ for the three most common particle species: protons, kaons and pions, and compare these with experimental data, as shown in Figure \ref{figure3}.  Since the modelled and experimental data are in a reasonable general agreement, we can now start mapping out the precise correspondence between the experiment and modelling data with variable initial conditions.   

\begin{figure}
\begin{center}
\includegraphics[width=6 in]{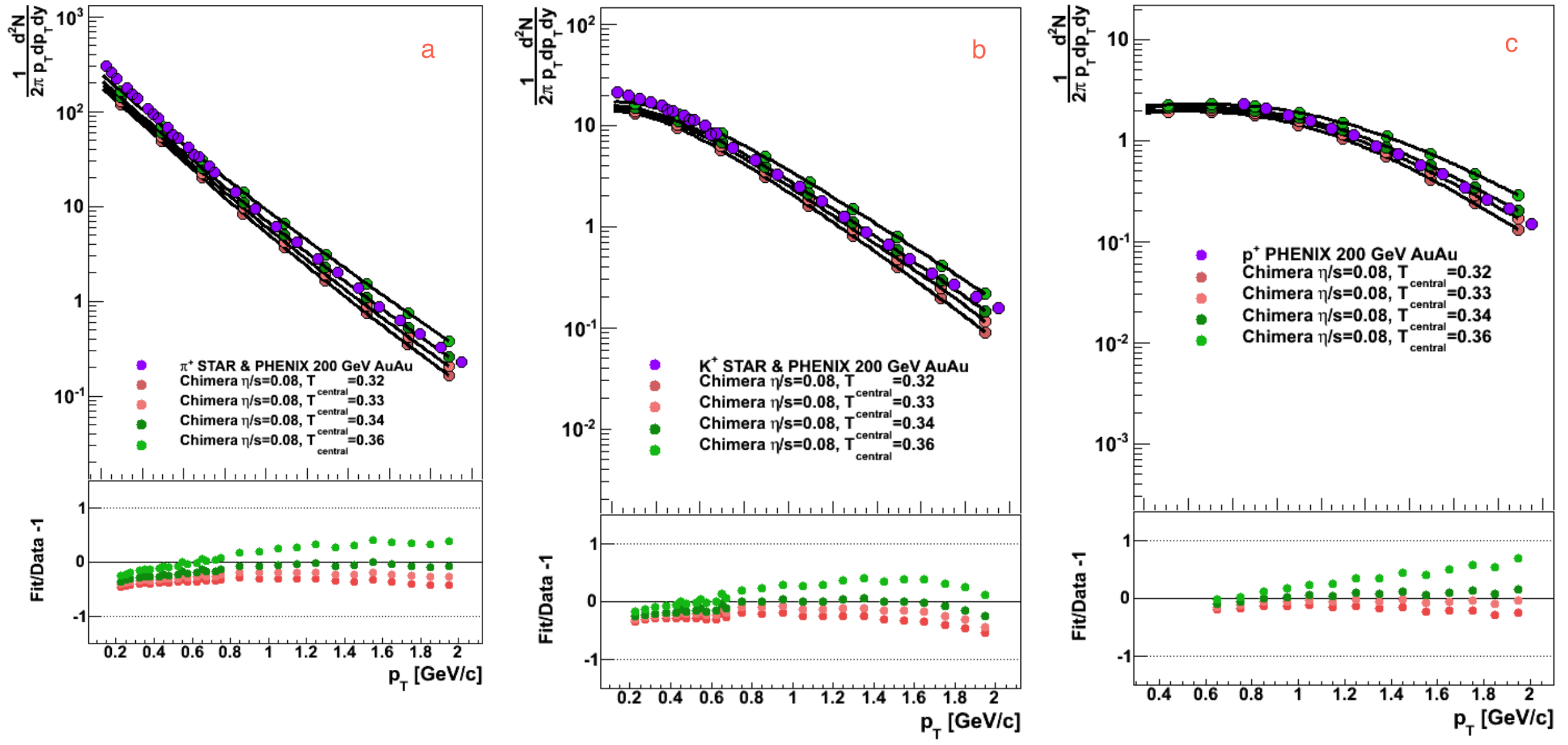}
\caption{\label{figure4} $\pi^+$ (panel $a$), K$^+$ (panel $b$), and $p$ (panel $c$) spectra distributions from CHIMERA and $\sqrt{s_{NN}}=200$ GeV 20\% central Au+Au RHIC data.  $\eta$/s is kept constant, while T$_{\mathrm{central}}$ is varied from 320 to 360 MeV.  Bottom sections of each panel show fractional agreement between data and a given set of parameters.  Model data are produced with \Ncoll\ scaling and pre-equilibrium flow off.}
\end{center}
\end{figure}

For each particle species, and for each of the three observables (HBT radii, spectra, and the elliptic flow coefficient), we obtain a set of data produced with parameters described above.  An example of particle spectra, plotted together with experimental data, is shown in Figure \ref{figure4}.  The figure illustrates a typical model-data comparison done to show the validity of hydrodynamical codes, and demonstrates a fraction of the available parameters generated by CHIMERA.  Note that each set of modelling data is fit using a Chebyshev polynomial (depending on the observable, the polynomial can have four or five parameters).

\begin{figure}
\begin{center}
\includegraphics[width=6 in]{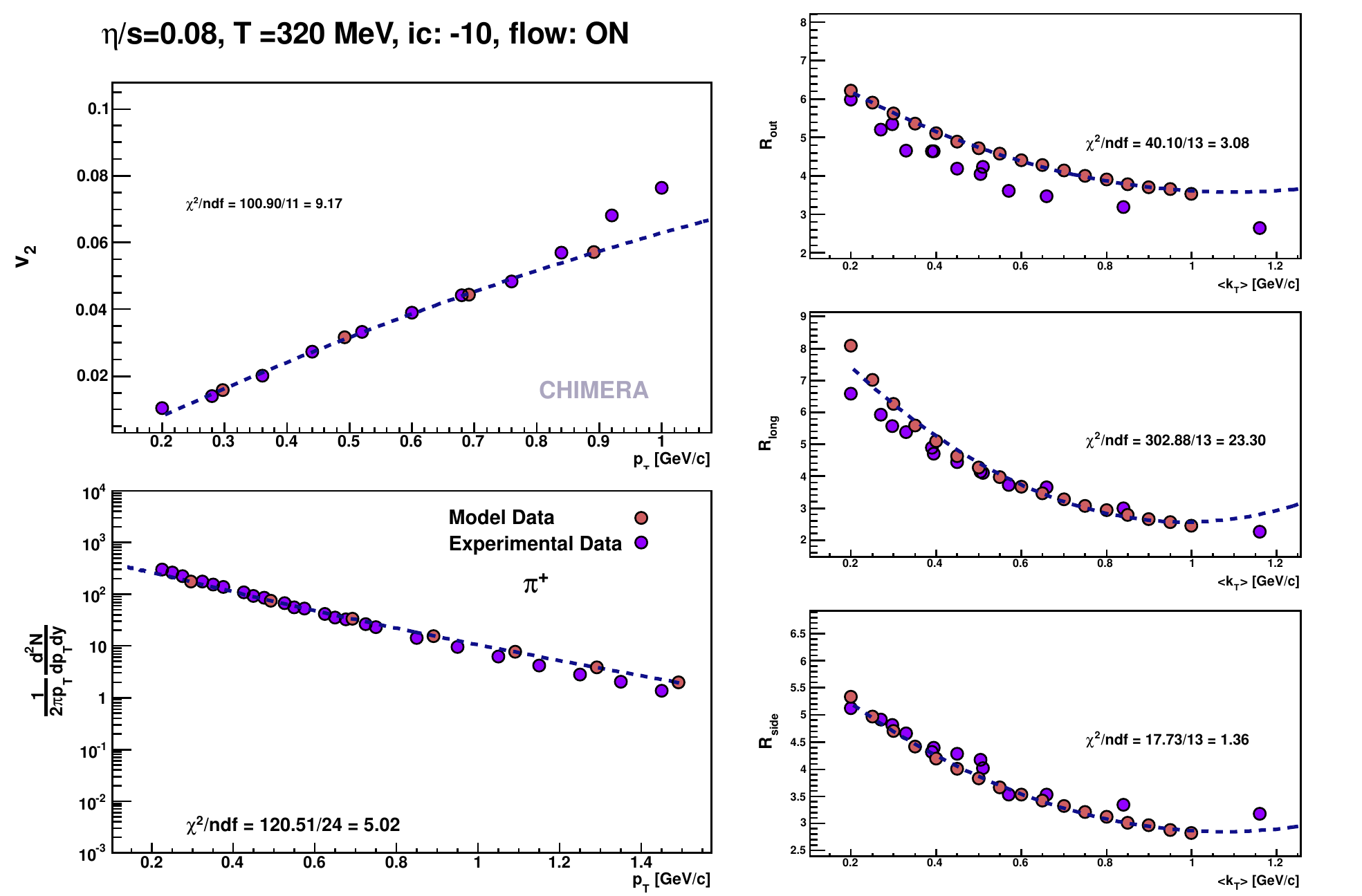}
\caption{\label{figure5} An example of $\chi^2$/ndf calculations for $\pi^+$ mesons. Each panel shows a comparison between model and experiment for each of the available quantities: $v_2$, spectra, and the three HBT radii (right side of the figure).  Pre-equilibrium flow is turned off, and the initial distribution scales as \Npart.  Experimental data is shown in purple, and modelling data points -- in salmon color filled circles.}
\end{center}
\end{figure}

For each experimental-modelling data pair, we calculate $\chi^2$, to quantify how well a given modelling curve describes a given set of experimental data points, as described in the previous section. Keeping all but two of the initial condition parameters constant, we are able to map out the $\chi^2$/ndf space for particle spectra, flow, and HBT radii, and then use the most probable overlap region to determine the most likely initial states parameter values.  In the study performed for this article, the $\chi^2$/ndf was mapped out for $\pi^+$ mesons, with experimental data obtained from both PHENIX \cite{Phenix_spectra}\cite{Phenix_flow} and STAR \cite{STAR_flow}\cite{STAR_hbt} collaborations.  An example, shown in Figure \ref{figure5}, shows a calculation of one map point for each $\chi^2$/ndf map (spectra, flow, and each of the HBT radii).  Figure \ref{figure5} calculations were performed with \Npart\ scaled modelling data, with pre-equilibrium flow turned on, at $\eta$/s=0.08, and T$_{\mathrm{central}}$=320 MeV.  By eye, these parameters represent the data fairly well, but exactly how well can be judged by the $\chi^2$/ndf shown in each panel of the figure.

\section{Results}
\begin{figure}
\begin{center}
\includegraphics[width=6.5 in]{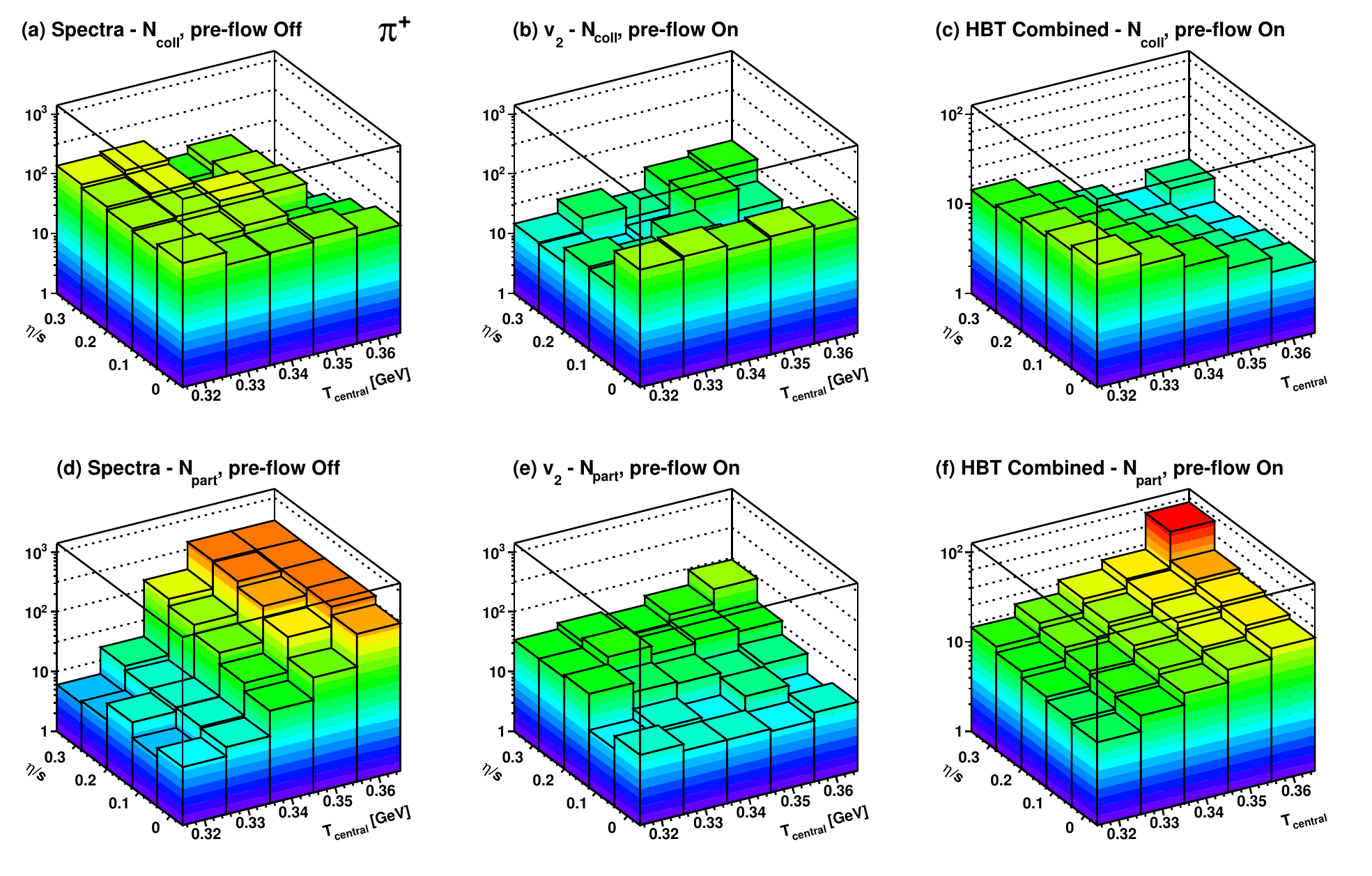}
\caption{\label{figure6}Comparison of $\chi^2$/ndf distributions for \Ncoll\ vs. \Npart\ scaling.  The top row has been produced with \Ncoll\ scaling, the bottom row scales with \Npart.  Panels $a$ and $d$ show $\chi^2$/ndf for $\pi^+$ spectra, $b$ and $e$ -- $\pi^+$ flow (v$_2$), and panels $c$ and $f$ show the distributions for the combined HBT radii. Pre-equilibrium flow was turned on for the four panels on the right, and turned off for the left-most panels.}
\end{center}
\end{figure}

The results of the study are summarized in Table \ref{table1} and Figure \ref{figure6}. Table \ref{table1} shows the minima from $\chi^2$/ndf distributions for all measurements currently available within CHIMERA made with $\pi^+$ mesons.  The $\chi^2$/ndf minima range from 4.25 for $\chi^2$/ndf calculations of the combined HBT radii with \Ncoll\ scaling and pre-equilibrium flow included, to 26.43 for the  $\chi^2$/ndf  distributions derived from calculations for v$_2$ without pre-flow and with \Npart\ scaling. Trends, evident from the table, are as follows: $\chi^2$/ndf  is significantly improved for v$_2$ and HBT measurements by turning on pre-equilibrium flow and by using \Ncoll\ scaling, while the reverse is true for pion spectra: the best $\chi^2$/ndf value is for \Npart\ scaling in simulations with pre-equilibrium flow turned off. 

\begin{table}
\caption{\label{table1}$\chi^2$/ndf minima for each of the three experimental observables: $\pi^+$ meson spectra, flow (v$_2$ coefficient), and the combined HBT radii.  The minima have been computed with pre-equilibrium flow turned on and off, and with initial state scaling as the number of binary collisions (\Ncoll) or number of participants (\Npart). }
\begin{center}
\begin{tabular}{lcccccc}
\br
 & \multicolumn{2}{c}{{\bf Spectra}} &\multicolumn{2}{c}{{\it{\bf v$_2$}}}&\multicolumn{2}{c}{{\bf HBT}}  \\ 
pre-flow &  on& off & on& off & on& off\\ 
 \mr
 {\bf \Ncoll} &18.73&11.86 &8.84  & 9.17 &  4.25& 12.06 \\ 
 {\bf \Npart} &5.02& 5.98 &9.17  & 26.43 &  8.46& 27.88 \\ 
\br 
\end{tabular}
\end{center}
\end{table} 

Figure \ref{figure6} shows $\chi^2$/ndf space distributions for spectra, v$_2$ and HBT radii of $\pi^+$ produced with \Ncoll\ scaling (panels $a$, $b$, and $c$) and \Npart\ scaling (panels $d$, $e$, $f$).  The distributions are drawn to display the most likely $\eta$/s and \Tcent\ values that would produce the experimental measurements.  A global $\eta$/s and \Tcent\ value would be at the intersection of the $\chi^2$/ndf minima for all three experimental quantities modelled under the same set of conditions.  However, this is not yet the case.  In \Ncoll-scaled data, $\eta$/s and \Tcent\ tend to higher values for spectra and to some extent for the HBT combined measurement as well, whereas for v$_2$ the most likely $\eta$/s is much lower ($\sim0.08$).  We also do not see a common minimum in the \Npart-scaled data: there the best $\eta$/s value for $\pi^+$ spectrum is 0.32, 0.08 for v$_2$, and 0.0001 for HBT.  The best \Tcent\ is the same for spectrum and HBT (320 MeV), but higher for v$_2$.  This indicates that the modelling part of CHIMERA is not yet completely physical.  

\section{Summary and outlook}

In summary, we presented an outline and the first tests of an evaluative framework for comparing hydrodynamical modelling attempts of the Quark Gluon Plasma to the experimental data available from the existing heavy ion experiments.  VH2 viscous hydrodynamical code was used as a base to generate modelling data used in this evaluation.  The data comparison was done for 10-20\% central $\sqrt{s_{NN}}=200$ GeV Au+Au collisions at RHIC, and aimed to study the effect of utilizing the initial scaling (\Ncoll\ vs \Npart).  In addition, we analysed possible effects of pre-equilibrium flow on the goodness of model's description of the data.  The result of the comparison finds that \Ncoll\ scaling might not be applicable when attempting to describe the spectra of the data derived from collision's bulk, but does a slightly better job than \Npart\ scaling at describing flow and the HBT variables.  We also found that pre-equilibrium flow significantly improves the goodness of fit for the HBT radii and v$_2$, as expected, but also undermined the model's description  of spectra. 

However, these are initial results, generated  primarily as proof of principle.  Currently, we plan on implementing a more realistic initial-state energy-density distribution, vary the EoS used for the analysis, and move on to three-dimensional hydro-dynamical modelling.  We are also performing $\tau_{\mathrm{start}}$ studies and studies to map out $\eta$/s dependence on temperature.  In addition, we are expanding our framework to analyse other collision centralities in Au+Au, other collision species and collision energies.  The onslaught of new data from RHIC and the LHC provides us with a unique and exciting opportunity for a comprehensive and detailed study of the matter produced in these collisions.  CHIMERA is an important tool for the Heavy Ion community, and we hope it would be used accordingly.

\ack
 The author thanks Scott Pratt for his invaluable contributions and discussions, Paul Romatschke and Matt Luzum for making their hydrodynamic code available, and Stephen Bass for the availability of UrQMD.  Most importantly, without Ron Soltz, Irakli Garishvili, Jason Newby, Andrew Glenn, Michael Cheng, and Alex Linden-Levy there would not be CHIMERA.   This work is supported by Lawrence Livermore National Laboratory  grant for laboratory-directed research and development, and has been assigned LLNL-PROC-488978.


\section*{References}

\begin{thebibliography}{25}
\bibitem{Perfect LiquidSTAR} Adams J {\it et al} (STAR Collaboration) 2005  {\it Nucl. Phys. A} {\bf 757} 102-83
\bibitem{Perfect_LiquidPHENIX} 	Adcox K {\it et al} (PHENIX ollaboration) 2005  {\it Nucl. Phys. A} {\bf 757} 184-283
\bibitem{Glauber} Alver B, Baker M, Loizides C and Steinberg P 2008 ({\it Preprint nucl-ex}/0805.4411)
\bibitem{CGC}Dumitru A 2011 {\it Nucl. Phy. A} {\bf 84} 71-5
\bibitem{HotQCD} Bazanov A {\it et al} (HotQCD Collaboration) 2009 {\it Phys. Rev. D} {\bf 80} 014504
\bibitem{Petrecky} Huovinen P and Petreczky P 2010 {\it Nucl. Phys. A} {\bf 837} 26-53
\bibitem{MITBag} Chodos A, Jaffe R L, Johnson K, Thorn C B and Weisskopf V F 1974 {\it Phys. Rev. D} {\bf 9} 3471
\bibitem{Hydro} Luzum M and Romatschke P 2008 {\it Phys. Rev. C} {\bf 78} 034915
\bibitem{Pratt} Vredevoogd J and Pratt S 2009 {\it Phys. Rev. C} {\bf 79} 044915
\bibitem{CooperFrye} Cooper F and Frye G 1974 {\it Phys. Rev. D} {\bf 10} 186-9
\bibitem{Bass} Bass S {\it et al} 1998 {\it Prog. Part. Nucl. Phys.} {\bf 41} 255-369
\bibitem{PhenixPi} Adare A {\it et al} (PHENIX Collaboration) 2008 {\it Phys. Rev. C} {\bf 77} 064907
\bibitem{Becattini} Becattini F 2002 {\it Nucl. Phys. A} {\bf 702} 336-40 
\bibitem{Phenix_spectra} Adler S S {\it et al} (PHENIX Collaboration) 2004 {\it Phys. Rev. C} {\bf 69} 034909-41
\bibitem{Phenix_flow}  Adler S S {\it et al} (PHENIX Collaboration) 2003 {\it Phys. Rev. Lett.} {\bf 91} 182301 
\bibitem{STAR_flow} Adams J {\it et al} (STAR Collaboration) 2005 {\it Phys. Rev. C} {\bf 72} 014904-27
\bibitem{STAR_hbt} Adams J {\it et al} (STAR Collaboration) 2005 {\it Phys. Rev. C} {\bf 71} 044906-27
\end{thebibliography}
\end{document}